\documentclass[12pt]{article}

\usepackage{amsmath,amssymb,amsfonts}
\usepackage{color}
\usepackage{graphicx}
\usepackage[footnotesize]{caption2}
\usepackage{bbm}
\usepackage{braket}
\usepackage[colorlinks]{hyperref}
\usepackage[figure,table]{hypcap}
\usepackage{bbold}
\usepackage{axodraw}
\usepackage{pstricks}

\newcommand{\GeV}{\:\text{GeV}}
\newcommand{\TeV}{\:\text{TeV}}

\newcommand{\eq}[1]{Eq.~\eqref{#1}}
\newcommand{\Figref}[1]{Fig.~\ref{#1}}
\newcommand{\Tabref}[1]{Tab.~\ref{#1}}

\def\msbar{\overline{\rm MS}}
\newcommand{\nn}{\nonumber}
\def\bea{\begin{eqnarray} }
\def\eea{ \end{eqnarray} } 

\allowdisplaybreaks[1]

\begin{document}

\begin{titlepage}

\begin{center}
{\Large\bf 
Flavour constraints on scenarios with two or three heavy squark generations\\
}

\vspace{1cm}
\renewcommand{\thefootnote}{\arabic{footnote}}

\textbf{
J\"orn Kersten\footnote[1]{Email: \texttt{joern.kersten@desy.de}}$^{(a)}$
and
Liliana Velasco-Sevilla\footnote[2]{Email: 
\texttt{liliana.velascosevilla@gmail.com}}$^{(a,b)}$
}
\\[5mm]
\textit{\small 
$^{(a)}$
University of Hamburg, II.\ Institute for Theoretical Physics,\\
Luruper Chaussee 149, 22761 Hamburg, Germany\\[2mm]
$^{(b)}$
 Instituto de F\'{\i}sica, Universidad Nacional Aut\'onoma de M\'exico,\\
 Apdo. Postal 20-364, 01000, M\'exico D.F., M\'exico.}
\end{center}

\vspace{1cm}

\begin{abstract}

\noindent 
We re-assess constraints from flavour-changing neutral currents in the kaon system on supersymmetric scenarios with a light gluino, two heavy generations of squarks and a lighter third generation.  We compute for
the first time limits in scenarios with three heavy squark families,
taking into account QCD corrections at the next-to-leading order.
We compare our limits with those in the case of two heavy families.
We use the mass insertion approximation and consider
contributions from gluino exchange to constrain the mixing between the
first and second squark generation.
While it is not possible to perform a general analysis,
we assess the relevance of each kind of
flavour- and CP-violating parameters. We also provide ready to use magic numbers for the computation of the Wilson coefficients at $2\GeV$ for these scenarios.

\end{abstract}

\end{titlepage}

\section{Introduction}

The initial searches for supersymmetric particles at the LHC indicate that  minimal supersymmetric scenarios where all supersymmetric particles have similar masses below a TeV are not realized in nature.
This, in addition to stringent bounds from supersymmetric contributions to flavour-changing neutral currents (FCNC), makes scenarios where some supersymmetric particles are significantly heavier than others more appealing, although these scenarios have been studied already for over a decade. With this motivation in mind and following the work in \cite{Kadota:2011cr}, we re-assess limits from FCNC in scenarios with two or three heavy families of squarks ($m_{\tilde q} \leq 10\TeV$), while keeping the gluino mass below $2\TeV$.
Studies of the QCD corrections in such setups, in particular for two
heavy families, were performed in \cite{Bagger:1997gg},
\cite{Contino:1998nw} and \cite{Barbieri:2010ar,Bertuzzo:2010un,Mescia:2012fg}.  The
first work considered leading-order (LO) QCD corrections and used the
vacuum insertion approximation (VIA) for the hadronic matrix elements.
The second one calculated next-to-LO (NLO) corrections and took into
account lattice results for the bag parameters appearing in the matrix
elements.  More recently, \cite{Barbieri:2010ar,Mescia:2012fg} discussed
specific patterns of family symmetry breaking with hierarchical squarks.
Besides, \cite{Barbieri:2010ar}
pointed out the need for a careful treatment of certain box diagrams in
some cases, which was studied in more detail in~\cite{Bertuzzo:2010un}.

The most dangerous supersymmetric contributions to FCNC often occur in
the kaon sector, in particular contributions to $\Delta m_K$ and the CP
violation parameter $\epsilon$.  Our motivations for revisiting the
corresponding limits are (i) significant progress in the determination
of the experimental values and theoretical expectations in the Standard
Model (SM) since the publication of \cite{Contino:1998nw}, (ii) the
availability of NLO matching conditions \cite{Ciuchini:2006dw,Virto:2009wm},
and (iii) the need for an NLO calculation of the QCD corrections in the
case of three heavy squark generations.

We proceed as follows: in section 2, we summarize, for the sake of
clarity, the way NLO QCD corrections are addressed for $\Delta S=2$
processes involving light gluinos and heavy scalars.  We provide
formulas that can be used to compute easily the renormalization group (RG)
evolution of the Wilson coefficients relevant for $\Delta S=2$
processes, both for the case of two heavy squark generations and for the
case of three heavy families.
We compare our results with the literature, finding that in some cases a
change of basis was missing in earlier work. 

In section 3, we re-assess limits from $\Delta m_K$ on the mass
$m_{\tilde q}$ of the heavy families of down-type squarks, restricting
ourselves to the contribution from gluino exchange.  We consider four
different combinations of non-zero flavour-violating parameters in the
mass insertion (MI) approximation, defined in \eq{eq:deltascases}.
A complete supersymmetric model of flavour should include a structure determining the form of Yukawa couplings, sfermion mass matrices and trilinear couplings, including their off-diagonal terms.  Since this is very model-dependent,  it is not possible to perform a general analysis of the limits on flavour-violating parameters. However, one can assess the relevance of each of them, or of combinations that appear naturally, such as the ans\"{a}tze of \eq{eq:deltascases}.
In section 3, we also comment on the difference between the scenarios
with three and two families of heavy scalars.  Finally, we determine
bounds from $\Delta m_K$ and $\epsilon$ on the real and imaginary parts
of flavour-violating parameters for given values of $m_{\tilde q}$ and
the mass of the gluino,~$m_{\tilde g}$.

\section{QCD corrections for heavy squarks and a light gluino \label{sec:qcdcorr}}
\subsection{Renormalization group evolution of the Wilson coefficients}

The effective Hamiltonian for $\Delta S=2$ transitions can be written as 
\begin{equation} \label{eq:effHam1}
H^{\Delta S=2} = \sum_{i=1}^5 C_i O_i + \sum_{i=1}^3 \tilde C_i \tilde O_i,
\end{equation}
where the operators $O_i$ are
\begin{eqnarray}
\label{eq:operatorsDF2}
&&O_1=\bar d^\alpha \gamma_\mu P_\text{L} s^\alpha \bar d^\beta \gamma^\mu P_\text{L}
s^\beta,\quad O_2=\bar d^\alpha P_\text{L} s^\alpha \bar d^\beta P_\text{L} s^\beta, 
\notag \\
&&O_3=\bar d^\alpha P_\text{L} s^\beta \bar d^\beta P_\text{L} s^\alpha,\quad \quad \quad
O_4=\bar d^\alpha P_\text{L} s^\alpha \bar d^\beta P_\text{R} s^\beta,  \notag \\
&&O_5= \bar d^\alpha P_\text{L} s^\beta \bar d^\beta P_\text{R} s^\alpha,
\end{eqnarray}
and $P_\text{L,R}$ are chirality projection operators,
$\alpha,\beta$ are color indices,
$\tilde O_i= O_i \, (\text{L}\leftrightarrow\text{R})$,
$\braket{\tilde O_i}=\braket{O_i}$, and
$\tilde C_i= C_i \, (\text{L}\leftrightarrow\text{R})$.
The RG evolution of the Wilson coefficients from a
high-energy scale $M$ to a lower energy $\mu$ is determined by the
$5\times 5$ evolution matrix $\widehat{W}[\mu,M]$,
\bea
\mathbf{C}(\mu)=\widehat{W}[\mu,M]\mathbf{C}(M),
\eea
where $\mathbf{C}$ represents a column vector with the $5$ components
$C_i$.  We denote row vectors with an arrow, such as the row vector
$\vec{\mathbf{O}}$ built up by the components $O_i$. At LO and NLO,
respectively, $\widehat{W}[\mu,M]$ can be expressed as
\cite{Ciuchini:1997bw}
\begin{eqnarray}
\widehat{W}[\mu,M]_\text{LO}&=&\widehat{U}[\mu,M] \nn\\
&=&\left[\frac{\alpha_{s}(M)} {\alpha_s(\mu)}\right]^{\widehat{\gamma}^{(0)T}\!/2\beta_0}, 
\nn\\
\widehat{W}[\mu,M]_\text{NLO}&=&
\widehat{U}[\mu,M] +
\frac{\alpha_s{(\mu)}}{4\pi} \widehat J(\tilde n_f) \widehat{U}[\mu,M] -
\frac{\alpha_s{(M)}}{4\pi} \widehat{U}[\mu,M] \widehat J(\tilde n_f),
\nonumber\\
\beta_0 & = & \frac{1}{3} \left( 11 N_c - 2 \tilde n_f \right),
\end{eqnarray}
where $N_c=3$ is the number of colors and $\tilde n_f$ equals the number
of active fermion flavors, $n_f$, at energies below the gluino mass,
$m_{\tilde g}$.  At higher energies,
\begin{equation} \label{eq:nftilde}
\tilde n_f = n_f + N_c + \frac{n_{\tilde q}}{4} ,
\end{equation}
where $n_{\tilde q}$ is the number of light squarks with a mass similar
to $m_{\tilde g}$ \cite{Contino:1998nw}.  The $5\times5$ matrices
$\widehat J$ were calculated in \cite{Ciuchini:1997bw}.
They depend on the renormalization scheme.  We use the LRI scheme
because it is also used in the lattice determination of the low-energy
matrix elements.
The one-loop anomalous dimension matrix (ADM), $\gamma^{(0)}$, is
scheme-independent.  It depends only on $N_c$ and is a $5\times 5$ matrix, since the five operators $O_i$ entering the Hamiltonian~\eqref{eq:effHam1} do not mix with others \cite{Ciuchini:1998ix} during the evolution down to low energies.

In \cite{Ciuchini:1997bw}, the ADMs and $\widehat J$ are given in the Fierz
basis of operators $O^+_i$, defined by
\bea
O_1&=&O^{+}_1,\nn\\
O_2&=&O^{+}_4,\nn\\
O_3&=&-\frac{1}{2}\left(O^{+}_4-\frac{1}{4} O^{+}_5\right),\nn\\
O_4&=&O^{+}_3,\nn\\
O_5&=&-\frac{1}{2}O^{+}_2.
\eea
Then we can define a matrix $V$ to transform between the basis of the
operators $O_i$ and the one of the operators $O^{+}_i$,
\begin{eqnarray}
\vec{\mathbf{O}}^{+} &=& \vec{\mathbf{O}} V^{-1},
\nonumber\\
\mathbf{C}^+ &=& V \mathbf{C} ,
\nonumber\\
V &=&
\left(
\begin{array}{ccccc}
1 & 0 & 0 &0 &0 \\
0 & 0 & 0  & -\frac{1}{2} & 0\\
0 & 0 & 0  & 1 & 0\\
0 & 1  & -\frac{1}{2} & 0 & 0\\
0 & 0  &  \frac{1}{8}  & 0 & 0
\end{array}
\right).
\label{eq:BasisChange}
\end{eqnarray}
We will denote all evolution matrices in the Fierz basis with a tilde
instead of a hat,
\bea
\widetilde{U}[\mu,M] &=& V^{-1}\widehat{U}[\mu,M] V,\nn\\
\widetilde{W}[\mu,M] &=& V^{-1}\widehat{W}[\mu,M] V.
\eea

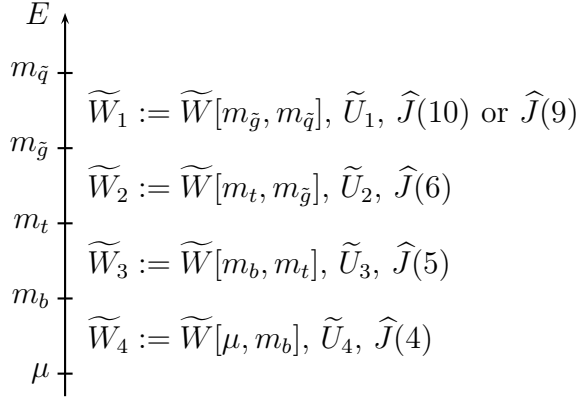
\begin{figure}
\centering
\begin{pspicture}(0,0.2)(7,5.5)
\psline{->}(0.2,0.2)(0.2,5.3)
\rput[r](0,5.3){$E$}
\rput[r](0,4.5){$m_{\tilde q}$}
\psline(0.1,4.5)(0.3,4.5)
\rput[l](0.5,4){$\widetilde W_1 := \widetilde W[m_{\tilde g},m_{\tilde q}]$, $\widetilde U_1$, $\widehat J(10)$ or $\widehat J(9)$}
\rput[r](0,3.5){$m_{\tilde g}$}
\psline(0.1,3.5)(0.3,3.5)
\rput[l](0.5,3){$\widetilde W_2 := \widetilde W[m_t,m_{\tilde g}]$, $\widetilde U_2$, $\widehat J(6)$}
\rput[r](0,2.5){$m_t$}
\psline(0.1,2.5)(0.3,2.5)
\rput[l](0.5,2){$\widetilde W_3 := \widetilde W[m_b,m_t]$, $\widetilde U_3$, $\widehat J(5)$}
\rput[r](0,1.5){$m_b$}
\psline(0.1,1.5)(0.3,1.5)
\rput[l](0.5,1){$\widetilde W_4 := \widetilde W[\mu,m_b]$, $\widetilde U_4$, $\widehat J(4)$}
\rput[r](0,0.5){$\mu$}
\psline(0.1,0.5)(0.3,0.5)
\end{pspicture}
\caption{Energy thresholds, evolution matrices and $\widehat J(\tilde n_f)$ involved in the RG evolution of the Wilson coefficients.
In the energy range between $m_{\tilde g}$ and $m_{\tilde q}$, $\widehat
J(10)$ and $\widehat J(9)$ refer to the case of $2$ and $3$ heavy squark
generations, respectively. In $\widetilde W_4$ we take $\mu=2\GeV$.
}
\label{fig:Scales}
\end{figure}

In the scenario under consideration, the RG evolution of the Wilson
coefficients starts at the mass scale of the heavy squarks,
$M=m_{\tilde q}$, and ends at $\mu=2\GeV$, where the matrix elements of
the $\Delta S=2$ operators are calculated.  Along the way, the gluinos and
heavy quarks have to be integrated out at their mass scales.  In the
case of two heavy squark generations, we assume the lighter squarks to
have the same mass as the gluino.%
\footnote{This also implies the absence of threshold corrections to
$\alpha_s$ at $m_{\tilde g}$.}
The corresponding mass scales are depicted in \Figref{fig:Scales}.
The figure also shows the matrices governing the evolution between the
scales.
The NLO evolution from $m_{\tilde{q}}$ down to $\mu$ yields
\begin{eqnarray}
\label{eq:evolutionsmatrix}
\mathbf{C}(\mu) &=& V^{-1} \widetilde W[\mu,m_{\tilde q}] V \mathbf{C}(m_{\tilde q}) ,
\nonumber\\
\widetilde W[\mu,m_{\tilde q}] &=& \left. \widetilde{W}[\mu,m_b] \widetilde{W}[m_b,m_t] \widetilde{W}[m_t,m_{\tilde{g}}] \widetilde{W}[m_{\tilde{g}},m_{\tilde{q}}] \right|_\text{NLO} .
\end{eqnarray}
The LO evolution is analogous, with the replacement
$\widetilde{W} \to \widetilde{U}$. 
The subscript ``NLO'' reflects the fact that we have to truncate the
expansion of the product of evolution matrices such that it contains
only terms up to $\mathcal{O}(\alpha_s)$~\cite{Buras:2001ra}.  Explicitly,
\bea
\widetilde{W}[\mu,m_{\tilde q}] &=& \widetilde U_4\widetilde U_3 \widetilde U_2 \widetilde U_1 +\widetilde U_4 \widetilde U_3 \widetilde U_2 (\widetilde W_1-\widetilde U_1) + \widetilde U_4 \widetilde U_3 (\widetilde W_2-\widetilde U_2)\widetilde U_1\nn\\
&& {} + \widetilde U_4 (\widetilde W_3-\widetilde U_3)\widetilde U_2 \widetilde U_1 + (\widetilde W_4 - \widetilde U_4)\widetilde  U_3 \widetilde U_2 \widetilde U_1.
\eea
$\widetilde{W}[\mu,m_{\tilde q}]$ depends on the values of $\alpha_s$
at the scales $m_{\tilde q}$, $m_{\tilde g}$, $m_t$, $m_b$ and $\mu$.
Plugging in the known values of $\alpha_s(m_t)$, $\alpha_s(m_b)$ and
$\alpha_s(\mu)$, we can write the evolution matrices
$\widetilde W[\mu,m_{\tilde g}]$ and
$\widetilde W[m_{\tilde g},m_{\tilde q}]$ as functions of the
free parameters $\alpha_s(m_{\tilde g})$ and $\alpha_s(m_{\tilde q})$
as well as the so-called magic numbers, analogously to Eqs.~(9) of
\cite{Contino:1998nw},
\begin{eqnarray}
\widetilde W[\mu,m_{\tilde g}] &=& \sum_{r=1}^5
\left[ b_\text{LO}^{(r)} + b_\text{NLO}^{(r)} + \eta_6 c^{(r)} \right]
\eta_6^{a_r} ,
\nonumber\\
\widetilde W[m_{\tilde g},m_{\tilde q}] &=& \sum_{r=1}^5
\left[ d^{(r)} + \eta_6 e^{(r)} + \eta_{\tilde g} \eta_6 f^{(r)} \right]
\eta_{\tilde g}^{a'_r} ,
\label{eq:MagicContino}
\end{eqnarray}
where
\begin{equation}
\eta_6 = \frac{\alpha_s(m_{\tilde g})}{\alpha_s(m_t)}
\quad,\quad
\eta_{\tilde g} = \frac{\alpha_s(m_{\tilde q})}{\alpha_s(m_{\tilde g})} ,
\end{equation}
and the magic numbers $b^{(r)}$, $c^{(r)}$, $d^{(r)}$, $e^{(r)}$ and
$f^{(r)}$ are $5\times5$ matrices.
However, as the product 
$\widetilde W[\mu,m_{\tilde g}] \widetilde W[m_{\tilde g},m_{\tilde q}]$
contains terms of order $\alpha_s^2$, to be consistent, we have to use the
single evolution matrix
\begin{equation} \label{eq:Magic}
\widetilde W[\mu,m_{\tilde q}] = \sum_{r,s} \left\{
\left[ b_\text{LO}^{(r)} + b_\text{NLO}^{(r)} + \eta_6 c^{(r)} \right]
 d^{(s)} +
b_\text{LO}^{(r)} \left[ e^{(s)} + \eta_{\tilde g} f^{(s)} \right] \eta_6
\right\} \eta_6^{a_r} \eta_{\tilde g}^{a'_s}
\end{equation}
rather than Eqs.~\eqref{eq:MagicContino}.
This approach is useful because for a given model, with three or two heavy
families of squarks, we have to calculate only $\eta_6$ and
$\eta_{\tilde g}$.  Together with the magic numbers, 
\eq{eq:Magic} then immediately yields
the values of the Wilson coefficients at $\mu=2\GeV$.

\subsection{Evolution between squark and gluino mass scales}
\label{ssec:3heavysq}

Due to the hierarchy between gluino and squark masses, integrating out all superparticles at the same scale would produce large logarithms. Therefore, we
proceed in two steps, first integrating out the heavy squark generations
at $m_{\tilde q}$ \cite{Bagger:1997gg,Contino:1998nw} and then the gluinos and light squarks at $m_{\tilde g}$.

\begin{figure}
\centering
\begin{picture}(480,200)(0,0)
\ArrowLine(10,175)(40,175)
\ArrowLine(10,125)(40,125)
\Vertex(40,175){2}
\ArrowLine(80,175)(110,175)
\Vertex(80,175){2}
\DashArrowLine(40,175)(80,175){3}
\DashArrowLine(40,125)(80,125){3}
\Gluon(40,175)(40,125){5}{5}
\Line(40,175)(40,125)
\Gluon(80,175)(80,125){-5}{5}
\Line(80,175)(80,125)
\Vertex(40,125){2}
\ArrowLine(80,125)(110,125)
\Vertex(80,125){2}
\Text(135,150)[1]{=}
\Text(60,110)[1]{(I)}
\ArrowLine(155,175)(195,175)
\ArrowLine(155,125)(195,125)
\Vertex(215,175){2}
\Vertex(215,125){2}
\DashArrowLine(195,175)(235,175){3}
\ArrowLine(235,175)(275,175)
\Vertex(195,175){2}
\Vertex(235,175){2}
\Vertex(195,125){2}
\Vertex(235,125){2}
\Gluon(195,175)(195,125){5}{5}
\Line(195,175)(195,125)
\Gluon(235,175)(235,125){-5}{5}
\Line(235,175)(235,125)
\DashArrowLine(195,125)(235,125){3}
\ArrowLine(235,125)(275,125)
\Vertex(235,125){2}
\Line(210,180)(220,170)
\Line(210,170)(220,180)
\Line(210,130)(220,120)
\Line(210,120)(220,130)
\Text(290,150)[1]{+}
\Text(215,110)[1]{(IIa)}
\ArrowLine(300,175)(340,175)
\ArrowLine(300,125)(340,125)
\DashArrowLine(340,175)(380,175){3}
\DashArrowLine(340,125)(380,125){3}
\Vertex(340,175){2}
\Vertex(380,175){2}
\Vertex(340,125){2}
\Vertex(380,125){2}
\Gluon(340,175)(340,125){5}{5}
\Line(340,175)(340,125)
\Gluon(380,175)(380,125){-5}{5}
\Line(380,175)(380,125)
\ArrowLine(380,175)(420,175)
\ArrowLine(380,125)(420,125)
\Line(349,180)(359,170)
\Line(349,170)(359,180)
\Line(362,180)(371,170)
\Line(362,170)(371,180)
\Line(355,130)(365,120)
\Line(355,120)(365,130)
\Text(360,110)[1]{(IIb)}
\Vertex(354,175){2}
\Vertex(367,175){2}
\Vertex(360,125){2}
\Text(431,150)[1]{+$\ \hdots$}
\ArrowLine(40,75)(80,75)
\Vertex(80,75){2}
\ArrowLine(120,75)(160,75)
\Vertex(120,75){2}
\DashArrowLine(80,75)(120,75){3}
\DashArrowLine(80,25)(120,25){3}
\Gluon(80,75)(80,25){5}{5}
\Line(80,75)(80,25)
\Gluon(120,75)(120,25){-5}{5}
\Line(120,75)(120,25)
\ArrowLine(40,25)(80,25)
\Vertex(80,25){2}
\ArrowLine(120,25)(160,25)
\Vertex(120,25){2}
\Line(95,80)(105,70)
\Line(95,70)(105,80)
\Line(95,30)(105,20)
\Line(95,20)(105,30)
\Vertex(100,75){2}
\Vertex(100,25){2}
\Text(180,50)[1]{=}
\Text(100,10)[1]{(IIa)}
\ArrowLine(200,57)(240,50)
\ArrowLine(240,50)(280,57)
\Vertex(240,50){2}
\ArrowLine(200,42)(240,50)
\ArrowLine(240,50)(280,42)
\Text(240,10)[1]{(IIIa)}
\Text(302,50)[1]{+}
\ArrowLine(320,75)(360,75)
\Vertex(360,75){2}
\ArrowLine(360,75)(400,75)
\GlueArc(360,50)(20,360,0){-5}{12}
\CArc(360,50)(18,0,360)
\ArrowLine(320,25)(360,25)
\Vertex(360,25){2}
\ArrowLine(360,25)(400,25)         
\Text(360,10)[1]{(IIIb)}
\Text(416,50)[1]{+$\ \hdots$}
\end{picture} 
\caption{Diagrams representing the process of integrating out the heavy particles. The series of diagrams in the upper line represents the transition from the complete $\Delta S=2$ box diagram to the mass insertion approximation, where the leading order is represented by diagram (IIa) and the next order by diagram (IIb).
Crosses denote mass insertions.
Diagrams (IIIa) and (IIIb) arise from diagram~(IIa) when integrating out the $d$-type squarks.}
\label{fig:effdiagrams}
\end{figure}
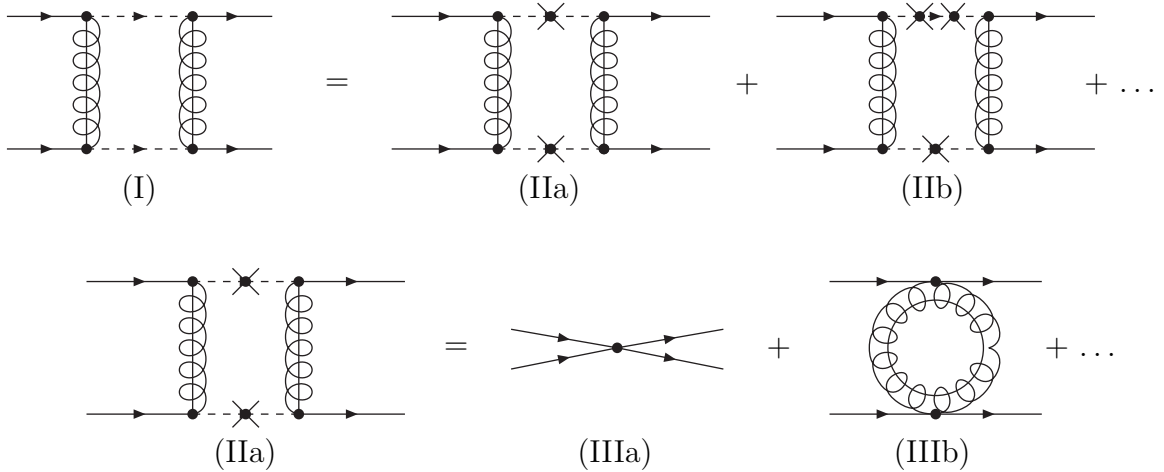

The LO matching at $m_{\tilde q}$ is visualized in \Figref{fig:effdiagrams}.
In the full theory, FCNC in the neutral kaon system stem from the
$\Delta S = 2$ box diagram~(I), where the scalar lines represent squark
mass eigenstates.  Using the MI approximation, we work
with flavour eigenstates and flavour-changing mass insertions,
represented by dashed lines and crosses in diagrams~(IIa,b), and
consider only diagram~(IIa).  This diagram contains only $\tilde d$ and
$\tilde s$ squarks.  For non-vanishing mixing between the first two and
the third generation, $\tilde b$ squarks appear in box diagrams with at
least three MI, i.e., in diagram~(IIb) and higher orders.  Such
contributions can be important, in particular for a light $\tilde b$.
As we aim to constrain the mixing between the first and the second
squark generation, we consider vanishing mixing with the third
generation.  Then the considered approximation is justified, and the
$\tilde b$ does not contribute to the matching.  Consequently, even in
the case of light third-genaration squarks, there are no contributions
from diagrams involving one heavy and one light squark, which require a
special treatment \cite{Barbieri:2010ar,Bertuzzo:2010un}.

In the effective theory below $m_{\tilde q}$ but above $m_{\tilde g}$,
diagrams~(IIIa) and (IIIb) of \Figref{fig:effdiagrams} yield FCNC\@.  The
former contains the $\Delta S = 2$ operators $O_i$ and $\tilde O_i$,
which are represented by the dot, and it is suppressed by $1/m^2_{\tilde q}$.
The latter diagram is suppressed by $m_{\tilde g}^2/m_{\tilde q}^4$ and
thus negligible \cite{Bagger:1997gg}.

Altogether, the final LO matching condition is (IIa)$=$(IIIa).  At NLO,
one has to consider diagrams with one more loop as well, which can be
found in \cite{Ciuchini:2006dw}.  The
resulting expressions for the gluino contributions to the Wilson
coefficients at high energy, $C_i^{\tilde g}(m_{\tilde q})$, are given in
Appendix~A of \cite{Ciuchini:2006dw} in terms of the MI parameters
$(\delta^d_{XY})_{12}$.%
{\footnote{We define
$(\delta^d_{XY})_{ij} :=
\frac{(m^2_{\tilde d\,XY})_{ij}^{}}%
{\sqrt{(m^2_{\tilde d\,XY})_{ii}^{}(m^2_{\tilde d\,XY})_{jj}^{}}}
$, where $X,Y \in\{$L,\,R$\}$ and where one has to use the soft mass
squared matrices in the super-CKM basis, where Yukawa couplings are
diagonal.
The corresponding expressions in the general case (without resorting to
the MI approximation) are given in~\cite{Virto:2009wm}.
}}
They are the same in both cases for the squark masses we consider.

The high-scale Wilson coefficients contain the strong
coupling and gluino mass at~$m_{\tilde q}$.
Therefore, we have to run these quantities up to this scale, using the
$\msbar$ (NDR) scheme as appropriate for the effective theory below
$m_{\tilde q}$.
We use the gluino pole mass $m_{\tilde g}$ as input parameter.
We transform to the running mass $M_3(m_{\tilde g})$, taking into account
one-loop corrections~\cite{Pierce:1996zz} involving gluinos and (if
applicable) one generation of squarks with mass $m_{\tilde g}$,
\begin{equation} \label{eq:GluinoMassConversion}
M_3(m_{\tilde g}) = m_{\tilde g} \left(
1 - \frac{15\,\alpha_s(m_{\tilde g})}{4\pi} +
\frac{n_{\tilde q}}{4} \frac{\alpha_s(m_{\tilde g})}{2\pi} \right) .
\end{equation}
In the case of three heavy generations of squarks, we run using the $\beta$ functions of \eq{eq:betafuncts_3heavysquarks}.  
For two heavy generations, the situation is more involved because the
$\beta$ functions for $\alpha_s$ and $M_3$ depend on the gluino-squark-quark
coupling, which differs from the gauge coupling in $\msbar$.
The corresponding $\beta$ functions and details of the calculation are given in
the appendix, Eqs.~\eqref{eq:betafuncts_2heavysquarks} and below.

As \cite{Ciuchini:2006dw} provides the results in the DRED
($\overline{\text{DR}}$) scheme, we have to convert $\alpha_s$ and $M_3$
to this scheme at $m_{\tilde q}$, using
\begin{eqnarray}
\alpha_s^{\rm{DRED}}(m_{\tilde q}) &=&
\alpha_s(m_{\tilde q}) \left( 1+\frac{\alpha_s(m_{\tilde q})}{4\pi} \right),
\nonumber\\
M_3^{\rm{DRED}}(m_{\tilde q}) &=&
M_3(m_{\tilde q}) \left( 1-3\,\frac{\alpha_s(m_{\tilde q})}{4\pi} \right) .
\label{eq:matching_alphamgls}
\end{eqnarray}

After performing the NLO matching, we compute the Wilson coefficients
$C(m_{\tilde q})$ in the RI scheme, using the translation matrix
$\Delta r^\text{DRED/RI}$ given in Eq.~(4.8) of \cite{Ciuchini:2006dw}.
Then we run these Wilson coefficients to $2\GeV$.

We stress that in the RI scheme it is not necessary to identify explicitly the effervescent operators in the intermediate steps before taking the limit $d\rightarrow 4$. When the limit $d\rightarrow 4$ is taken, the RI scheme guarantees that the expressions for the five physical operators $O_i$ contain all possible contributions, including those given by the effervescent operators, without the need to identify them \cite{Ciuchini:2006dw}.%
\footnote{We follow the definition of LO and NLO Wilson coefficients from  Eq.~(2.5) of \cite{Ciuchini:2006dw}.}

From $m_{\tilde q}$ down to $m_{\tilde g}$, gluinos and (if present)
light squarks influence the RG evolution of the Wilson coefficients by
changing the running of $\alpha_s$ and the two-loop ADM of the
$\Delta S = 2$ operators.  The impact on the ADM stems only from loops
with gluinos and light squarks in the gluon propagator.  Consequently,
it can be taken into account by using $\widehat J(n_f)$ of
\cite{Ciuchini:1997bw}, where only loops with fermions and gluons were
considered, and replacing $n_f$ by $\tilde n_f$ \cite{Contino:1998nw},
as mentioned above.  Explicitly, $\tilde n_f = 9$ if all squarks are
heavy.  If only two squark generations are heavy, $n_{\tilde q} = 4$
squarks are light and thus $\tilde n_f = 10$.
Using these considerations and the appropriate formulas from the
previous subsection, we calculated the magic numbers determining the NLO
QCD corrections in both cases.  We list them in
Appendix~\ref{eq:magicnumbers}.  Our results differ from those of
\cite{Contino:1998nw}.  As the disagreement in the magic numbers $d$,
$e$ and $f$ vanishes for two heavy squark generations if we set
$V=\mathbbm{1}$, we conclude that \cite{Contino:1998nw} did not  take
into account the change of basis \eqref{eq:BasisChange}.

After the gluino and possible light squarks decouple at $m_{\tilde g}$,
the system behaves as in
\cite{Ciuchini:1997bw,Ciuchini:1998ix,Buras:2001ra}.  Therefore, the
magic numbers $b$ and $c$ should equal those given in
\cite{Ciuchini:1998ix}, up to a different ordering of the components $r$
and small differences due to changes in the experimental input parameters.
We find larger differences than expected for some of the magic numbers.
The main reason seems to be that $\mathcal{O}(\alpha_s^2)$ terms were
not discarded in \cite{Ciuchini:1998ix}.

\section{Limits on flavour- and CP-violating parameters from $\Delta m_K$ and $\epsilon$}

\subsection{Limits from $\Delta m_K$}

In order to set a limit on the supersymmetric contribution to $\Delta m_K$, we write
\begin{equation}
\Delta m_K^{\text{T}}=\Delta m_K^{\text{SM}} +\Delta m_K^{\tilde g},
\end{equation}
where $\Delta m_K^{\text{T}}$ represents the total theoretical value of $\Delta m_K$, while $ \Delta m_K^{\text{SM}}$ and $\Delta m_K^{\tilde g}$, respectively, represent the SM and the supersymmetric contributions due to $\tilde g$-$\tilde q$ diagrams.  Even in the SM, precise computations of  $ \Delta m_K^{\text{SM}}$  are not possible due to unknown long-distance contributions \cite{Donoghue:1992dd}.  Hence, the best one can do is to compare the best estimate obtained from the short-distance contributions \cite{Brod:2011ty}, denoted by $\Delta m_K^{\text{SD}}$, to the experimental value 
\bea
\Delta m_K^\text{exp}&=&(3.483  \pm 0.0059)\times 10^{-15} \GeV,\nn\\
\Delta m_K^\text{SD}&=&(3.1 \pm 1.2)\times 10^{-15} \GeV,
\label{eq:Dmkcentral}
\eea
where all quoted errors correspond to the $1\,\sigma$ C.L\@.
$\Delta m_K^{\text{SD}}$ was obtained by taking into account NNLO contributions from the charm quark. We can see that its central value already accounts for 86\,\% of the experimental central value, but its uncertainty can easily account for the reported experimental value within the $1\,\sigma$ C.L\@. Given the lack of information on long-distance contributions, the best we can do to extract limits in extensions of the SM is to use $\Delta m_K^\text{exp}-\Delta m_K^\text{SD}$ as a constraint on the \emph{order of magnitude} of $\Delta m_K^{\tilde g}$. In short, it cannot exceed $10^{-15}\GeV$, but lower limits based on whether or not $\Delta m_K^\text{SD} + \Delta m_K^{\tilde g}$ can actually be as large as the experimental value cannot be obtained.  Furthermore, in some cases the importance of NLO corrections cannot be really appreciated because the differences between LO and NLO lie in the range of some units of $10^{-15}\GeV$.  Therefore, we have to be cautious  when using this for setting bounds in models for physics beyond the SM\@. As in the case of $\epsilon'$, there are many models that can pretty easily yield a change of $\Delta m_K$ by some units of $10^{-15}\GeV$, so meaningful constraints can indeed be obtained. However, we have to include a generous consideration of all the possible uncertainties in the calculation of $\Delta m_K^\text{SD} = 2\,\text{Re}\braket{  K^0|H_\text{SM}^{\Delta S=2} |\overline{K}^0}$.  The form of this theoretical expression at the known perturbative QCD level can be found in \cite{Brod:2011ty}.
For the gluino contribution,  we have 
\bea
\label{eq:DeltamK}
&&\Delta m_K^{\tilde g}= 2\,\text{Re}\braket{K^0|   H_{\tilde g}^{\Delta S=2}     |\overline{K}^0}= 2\,\text{Re}\braket{K^0|\sum_{i=1}^{5} C^{\tilde g}_i O_i +  \sum_{i=1}^{3} \tilde C^{\tilde g}_i \tilde O_i|\overline{K}^0}.
\eea
The Wilson coefficients at $\mu=2\GeV$ are determined by \eq{eq:Magic}. The matrix elements of the operators are \cite{Ciuchini:1998ix}
\bea
\label{eq:defmatrelementops}
\braket{ K^0|  O_1 |\overline{K}^0}&=& \frac{1}{3} M_K f^2_K B_1(\mu),\nn\\
\braket{ K^0|  O_i |\overline{K}^0}&=& k_i \left( \frac{M_K}{m_s(\mu)+m_d(\mu)} \right)^2 M_K f^2_K B_i(\mu),\quad i=2, 3, 4, 5, \nn\\
k_i&=& \frac{1}{8}\left(-\tfrac{5}{3}, \tfrac{1}{3},2,\tfrac{2}{3}\right)_i.
\eea
These expressions differ by a factor of $\frac{1}{8 M_K}$ from the
corresponding ones in the lattice study \cite{Bertone:2012cu}, whose
results for the bag parameters $B_i$ we use.
A factor $\frac{1}{4}$ stems from a different definition of the
operators and a factor $\frac{1}{2M_K}$ from the normalization of the
kaon states.%
\footnote{Cf.\ Eq.~(5) of \cite{Babich:2006bh},
$\Delta m_K = 2\,\text{Re}\braket{K^0|H^{\Delta S=2}|\overline{K}^0}/M_K$.
This work uses the same state normalization as \cite{Bertone:2012cu},
as one can verify by comparing their definitions of operators and matrix
elements, taking into account that the kaon decay constant is normalized
differently ($f_K$ of \cite{Bertone:2012cu} corresponds to
$\sqrt{2} F_K$ of~\cite{Babich:2006bh}).}
The definition \eqref{eq:defmatrelementops} ensures that the numerical
values of the bag parameters are the same in both conventions.
The VIA corresponds to
\bea
\label{eq:defmatrelementsVIA}
\braket{ K^0|  O_1 |\overline{K}^0}_{\text{VIA}}&=& \frac{1}{3} M_K f^2_K ,\nn\\
\braket{ K^0|  O_i |\overline{K}^0}_{\text{VIA}}&=& k_i \left( \frac{M_K}{m_s(\mu)+m_d(\mu)} \right)^2 M_K f^2_K,\quad  i=2, 3,
\nonumber\\
\braket{ K^0|  O_4 |\overline{K}^0}_{\text{VIA}}&=&\left[ \frac{1}{24}+ \frac{1}{4}\left(\frac{M_K}{ m_s(\mu)+m_d(\mu)  } \right)^2 \right] M_K f^2_K,\nn\\
\braket{ K^0|  O_5 |\overline{K}^0}_{\text{VIA}}&=&\left[ \frac{1}{8}+ \frac{1}{12}\left(\frac{M_K}{ m_s(\mu)+m_d(\mu)  } \right)^2 \right]  M_K f^2_K.
\eea
In Appendix~\ref{app:exp} we give the values of the experimental and the lattice parameters we use.

With the above points in mind, once we have the value of  $\Delta m_K^{\text{SD}}$, its uncertainty, that we call $\sigma_{\Delta m_K^{\text{SD}}}$, the experimental value and its uncertainty, $\sigma_{\Delta m_K^{\text{exp}}}$, we can set a limit on the supersymmetric contribution, $\Delta m_K^{\tilde g}$, using
\bea
\label{eq:limitDmk}
|\Delta m_K^{\tilde g} |< \left| \overline{\Delta m_K^{\text{exp}}} - \overline{ \Delta m_K^{\text{SM}}} + 2 \, (\sigma_{\Delta m_K^{\text{SD}}}+  \sigma_{\Delta m_K^{\text{exp}}}) \right|=2.8 \times 10^{-15}\GeV,
\eea
where $\overline{ \Delta m_K^{\text{SD}}}$ and $\overline{ \Delta m_K^{\text{exp}}}$ denote the central values.

\subsection{Limits from $\epsilon$}

As is also well-known, we can extract limits on the imaginary parts of some combinations of the parameters $(\delta_{XY}^d)_{12}$ using the CP-violating parameter $\epsilon$.  Currently, its experimental and SM \cite{Brod:2011ty} values are, respectively,
\bea
|\epsilon|&=& (2.228 \pm 0.011)\times 10^{-3},\nonumber \\
|\epsilon^\text{NNLO}|&=& (1.81 \pm 0.28 ) \times 10^{-3}.
\eea
The gluino contribution is
\begin{equation}
|\epsilon^{\tilde g}|=\kappa_\epsilon  \frac{ |{\rm{Im}} \braket{  K^0|H_{\tilde g}^{\Delta S=2} |\overline{K}^0}|}
{\sqrt{2} \, \Delta m_K},
\end{equation}
with $H_{\tilde g}^{\Delta S=2}$ as given in \eq{eq:DeltamK} and a
correction factor $\kappa_\epsilon$ \cite{Buras:2008nn}.  We use
the same value $\kappa_\epsilon=0.923$ \cite{Blum:2011ng} as \cite{Brod:2011ty}.
The limits are obtained using a condition analogous to \eq{eq:limitDmk},
with the obvious replacements, yielding
 \bea
 |\epsilon^{\tilde g}| < 1.0 \times 10^{-3}.
 \eea

\subsection{Scenarios considered}

As mentioned, we consider two scenarios for the sfermion mass
spectrum.  The first is Effective SUSY \cite{Cohen:1996vb}, where two generations of squarks remain heavy, while the other decouples at a lower scale, together with the gluino.  The second is a scenario where all three families of squarks are considerably heavier than the gluino, which occurs, for example, in the $G_2$-MSSM \cite{Acharya:2008zi}. 
We compare for each of these scenarios limits on the squark masses from $\Delta m_K$ in order to assess both the importance of the QCD corrections and the importance of the treatment described in section~\ref{ssec:3heavysq}.
Afterwards we extract limits on MI parameters for the scenario of three
heavy families. A study for completely general values
of the MI parameters is beyond the scope of this work, because
even in our constrained scenario where only the first- and
second-generation squarks mix, there are contributions from the four
parameters $(\delta^d_\text{LL})_{12}$, $(\delta^d_\text{RR})_{12}$, 
$(\delta^d_\text{LR})_{12}$ and $(\delta^d_\text{RL})_{12}$, 
containing six different combinations.
Instead, in order to assess the relevance of each kind of parameter, we
consider four cases that appear often in models,
\begin{align}
\label{eq:deltascases}
\text{I}&: \quad (\delta^d_\text{LL})_{12}=K,\ (\delta^d_\text{LR})_{12}=(\delta^d_\text{RL})_{12}=(\delta^d_\text{RR})_{12}=0,\nn\\
\text{II}&: \quad (\delta^d_\text{LR})_{12}=K,\ (\delta^d_\text{LL})_{12}=(\delta^d_\text{RL})_{12}=(\delta^d_\text{RR})_{12}=0,\nn\\
\text{III}&: \quad (\delta^d_\text{LL})_{12}=(\delta^d_\text{RR})_{12}=K,\ (\delta^d_\text{LR})_{12}=(\delta^d_\text{RL})_{12}=0, \nonumber\\
\text{IV}&: \quad (\delta^d_\text{LR})_{12}=(\delta^d_\text{RL})_{12}=K,\ (\delta^d_\text{LL})_{12}=(\delta^d_\text{RR})_{12}=0,
\end{align}
which also allows us to compare directly to \cite{Contino:1998nw}.

\subsection{Numerical results}

\begin{figure}
\centering
\includegraphics{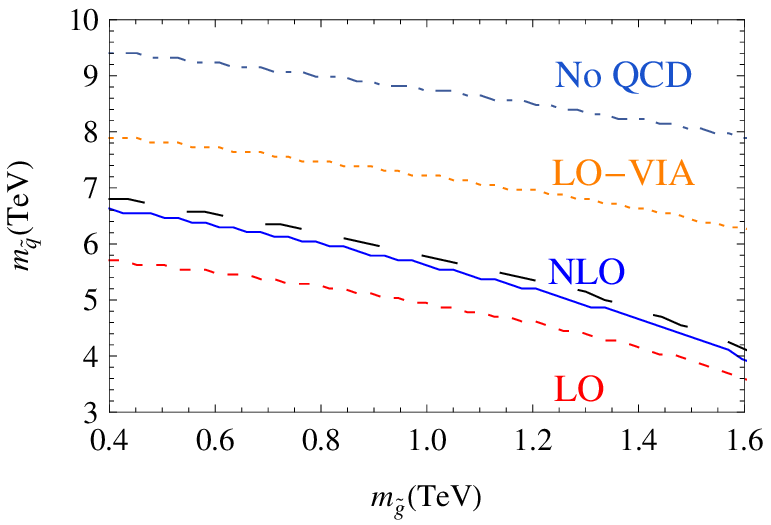}
\includegraphics{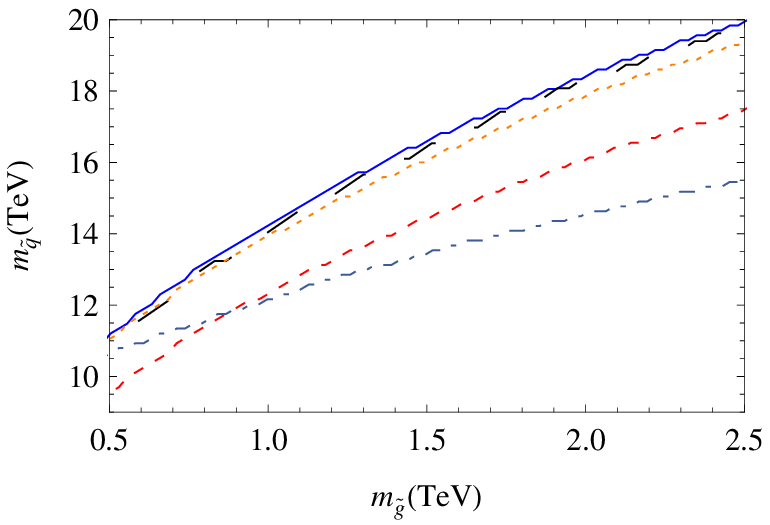}\\[2mm]
\includegraphics{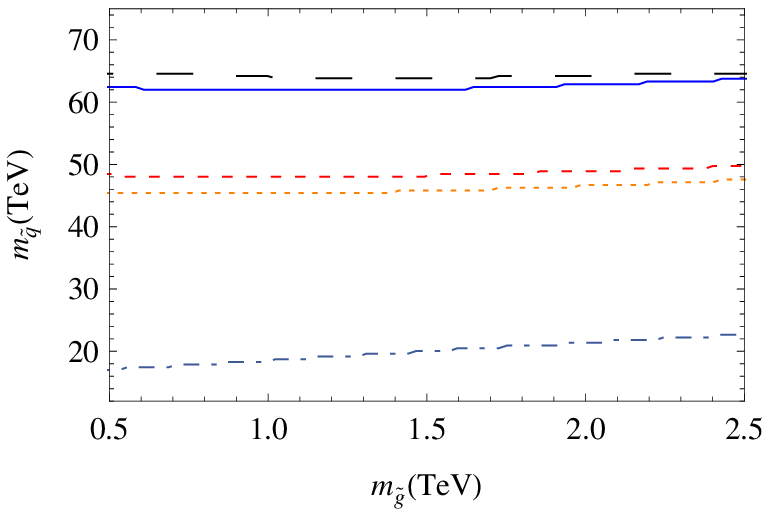}
\includegraphics{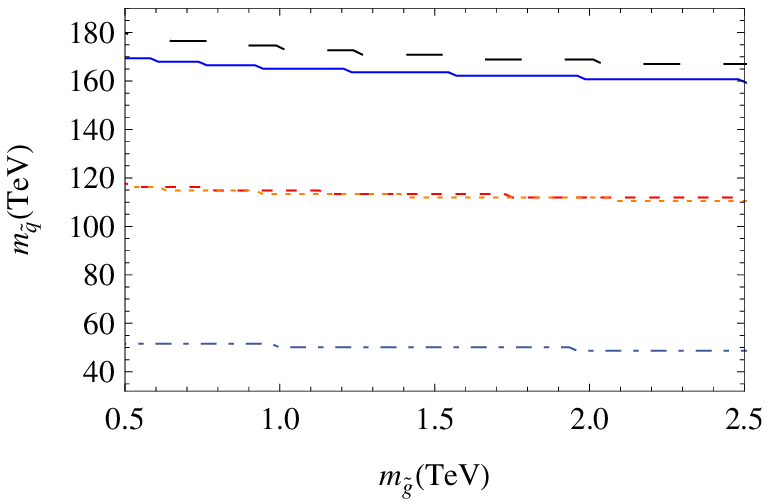}
\caption{Lower limits on the squark masses for the different cases of \eq{eq:deltascases} with $K=0.22$, derived from \eq{eq:limitDmk}.  The blue continuous line corresponds to the NLO evolution, the medium-dashed (red) line corresponds to the LO approximation, the dotted (orange) line corresponds to the  LO  evolution using the VIA for the bag parameters and the dot-dashed (light blue) line to the results without QCD corrections. We have indicated the labels just for  case I, but for the rest of the cases the coding is the same. For comparison, we have additionally plotted a long-dashed black line which represents the NLO evolution for the case where only two squark families are heavy.
}
\label{fig:cases3h}
\end{figure}

\paragraph{Limits on the squark masses}
In \Figref{fig:cases3h}, we present the lower limits on the heavy squark masses $m_{\tilde q}$ according to \eq{eq:limitDmk} as a function of the mass of the  gluino $m_{\tilde g}$,  for a fixed value of the flavour parameter $K=0.22$. The plots in \Figref{fig:cases3h} serve mainly (i) for comparison to Figs.~1 and 2 of \cite{Contino:1998nw}{\footnote{This is why we also plotted values of $m_{\tilde g}$ lower than the LHC bound around $900\GeV$. }}, (ii) to re-assess the importance of the NLO QCD corrections and (iii) to understand the difference between the scenarios with two and three heavy families of squarks.

We can see that the latter difference 
is not significant. In principle, this justifies using the QCD corrections of \cite{Contino:1998nw} also for scenarios with three heavy squark families, as done in \cite{Kadota:2011cr} as a first approximation. However, as mentioned above, a change of basis was missing in~\cite{Contino:1998nw}. Therefore, the limits shown in \Figref{fig:cases3h} are different from those in~\cite{Contino:1998nw}. Nevertheless, the qualitative behavior and order of magnitude of the limits are the same.
Regarding the QCD corrections, each step of improving the accuracy
(considering LO QCD corrections with VIA, including lattice bag
parameters and finally taking into account NLO QCD corrections)
can have a drastic effect.  Only case II may be considered an exception,
since here some of the corrections happen to cancel partially.  We can
also see that NLO QCD corrections, as opposed to just LO corrections,
are indeed relevant, especially for cases III and IV. 

We note that using different values of $K$ can lower
or raise considerably the limit on~$m_{\tilde q}$. For example, for
case~III and $K=0.1$, the limit becomes roughly
$m_{\tilde q} > 29\TeV$~\cite{VelascoSevilla:2012dy}, as opposed to $m_{\tilde q} > 62 \TeV$ for $K=0.22$, for $m_{\tilde g}=900 \GeV$.

\begin{figure}
\includegraphics{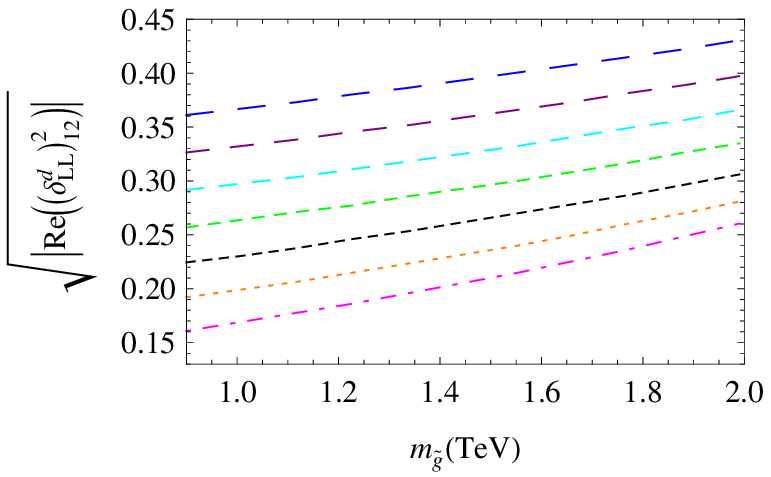}
\includegraphics{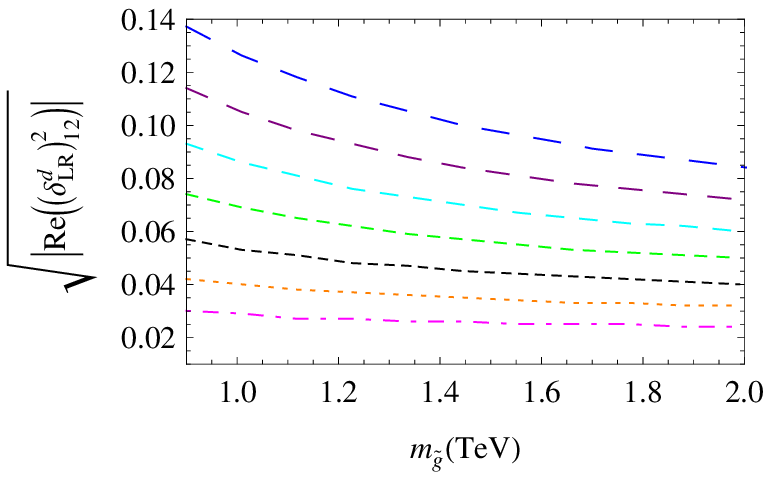}\\[2mm]
\includegraphics{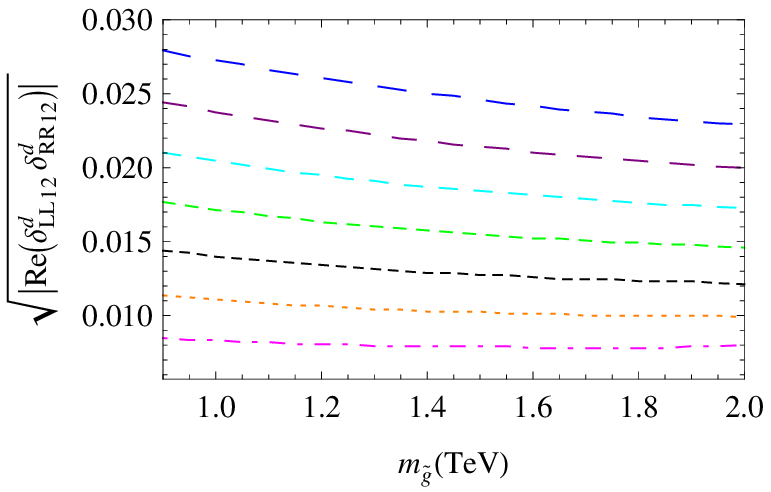}
\includegraphics{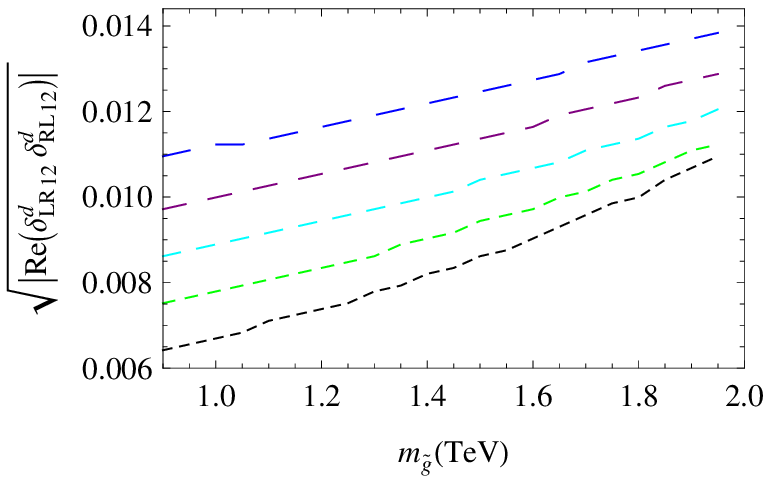}
\caption{Upper limits on $\sqrt{ |{\rm{Re}} ((\delta^d_{XY})_{12} (\delta^d_{X'Y'})_{12} )|}$, derived using $\Delta m_K$.  In all cases only the flavour-violating parameters shown on the vertical axis have been set to non-zero values. The limits on  $\sqrt{|{\rm{Re}}(\delta^d_{\text{RR}})^2_{12}|}$ and $\sqrt{|{\rm{Re}}(\delta^d_{\text{RL}})^2_{12}|}$ can be obtained by interchanging $\text{R}\leftrightarrow\text{L}$ in the upper two panels. The two lower panels show, respectively, the cases, where $\sqrt{|{\rm{Re}}((\delta^d_{\text{LL}})_{12} (\delta^d_{\text{RR}})_{12} ) |}$ or $\sqrt{|{\rm{Re}}((\delta^d_{\text{LR}})_{12} (\delta^d_{\text{RL}})_{12}) |}$ are not zero.  The  different curves correspond, from bottom to top, to  $m_{\tilde q}=4,\,\dots,\,10\TeV$, except for case IV, where we plot only from $m_{\tilde q}=6 \TeV$ up to $m_{\tilde q}=10 \TeV$.
\label{fig:Redeltas}}
\end{figure}

\paragraph{Limits on MI parameters from ${\Delta m_K}$}
Having found only small differences between the cases with two and
three heavy squark families, we will restrict ourselves to the latter
scenario in the following.
In \Figref{fig:Redeltas} we present curves for  fixed values of $m_{\tilde q}$, plotting $m_{\tilde g}$ against the upper limits on $\sqrt{|{\rm{Re}}((\delta^d_{XY})_{12} (\delta^d_{X'Y'})_{12} ) |}$ determined by requiring a sufficiently small gluino contribution to $\Delta m_K$, \eq{eq:limitDmk}.
The different curves correspond, from bottom to top, to  $m_{\tilde q}=4,\,\dots,\,10\TeV$, except for case IV, where we plot only from $m_{\tilde q}=6 \TeV$ up to $m_{\tilde q}=10 \TeV$.  For $m_{\tilde q}=2\TeV$ and the maximum value of the gluino mass we use, $m_{\tilde g}=2\TeV$, the condition $m_{\tilde q} \gg m_{\tilde g}$ defining the scenario under consideration would be violated.  For case IV, this approximation yields reliable results only for $m_{\tilde q}\geq 6 \TeV$.
The use of $\Delta m_K$ indeed yields relevant limits on all MI
parameters $(\delta^d_{XY})_{12}$. 
They turn out to be strongest if Re$(\delta^d_\text{LR})_{12}$ or Re$(\delta^d_\text{RL})_{12}$ are non-zero. 
For cases when only the real part of one parameter is allowed to be
non-zero and for gluino masses above $900\GeV$, the limit is typically
of order $10^{-1}$ and never  below $0.02$. Of course, the lower $m_{\tilde q}$, the lower the limit.
For the cases where either $\sqrt{|{\rm{Re}}((\delta^d_{\text{LL}})_{12} (\delta^d_{\text{RR}})_{12} ) |}$ or $\sqrt{|{\rm{Re}}((\delta^d_{\text{LR}})_{12} (\delta^d_{\text{RL}})_{12} ) |}$ are non-zero,  the limits on the combinations are of $\mathcal{O}(10^{-2})$ or even lower for case IV.

\begin{figure}
\includegraphics{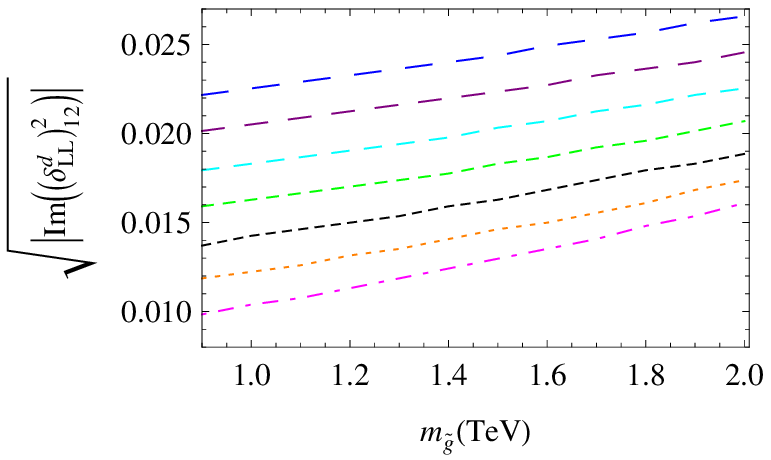}
\includegraphics{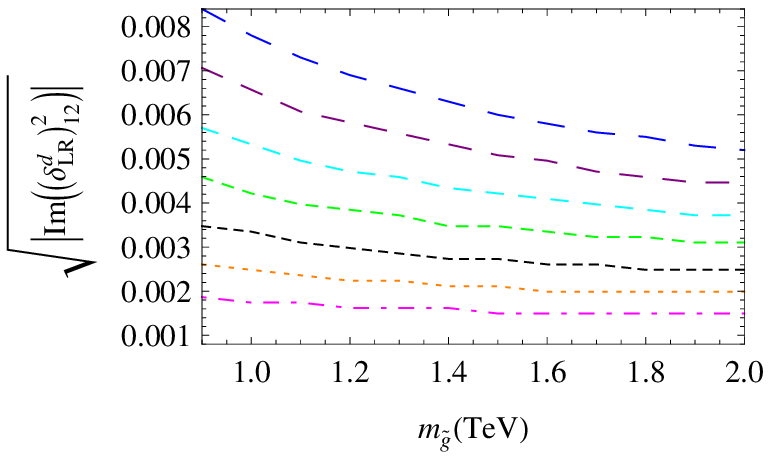}\\[2mm]
\includegraphics{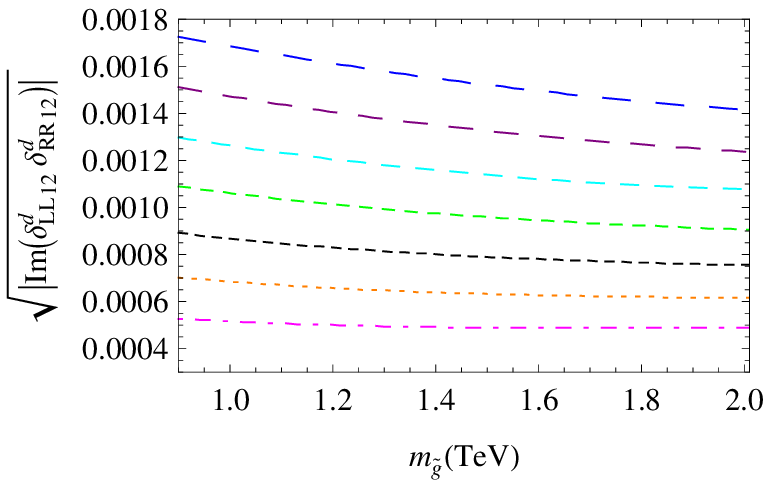}
\includegraphics{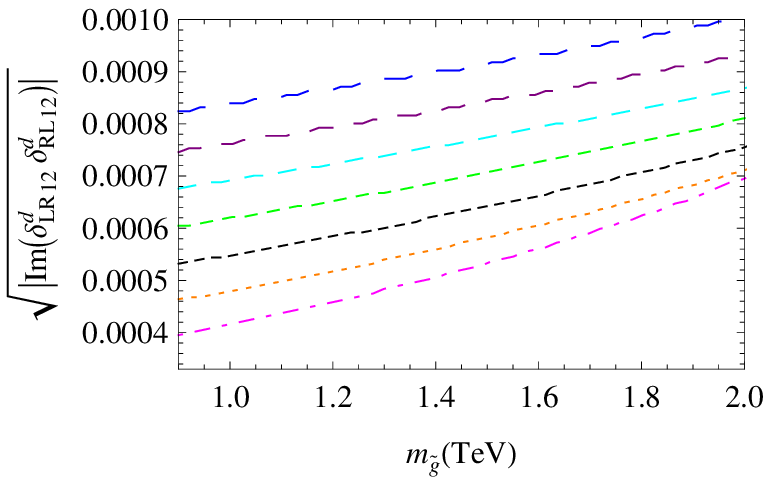}
\caption{Upper limits on  $\sqrt{|\text{Im}((\delta^d_{XY})_{12} (\delta^d_{X'Y'})_{12})|}$.  The limits on $\sqrt{|\text{Im}(\delta^d_{\text{RR}})^{2}_{12}|}$  and $\sqrt{|\text{Im}(\delta^d_{\text{RL}})^{2}_{12}|}$ can be obtained by interchanging $\text{R}\leftrightarrow\text{L}$ in the corresponding upper two panels. The lower two panels
correspond to the cases where  only the combinations $\sqrt{|\text{Im}((\delta^d_{\text{LL}})_{12} (\delta^d_{\text{RR}})_{12}) |}$ or $\sqrt{|\text{Im}((\delta^d_{\text{LR}})_{12} (\delta^d_{\text{RL}})_{12}) |}$, respectively, are non-zero.
The  different curves correspond, from bottom to top, to $m_{\tilde q}=4,\,\dots, 10 \TeV$. 
\label{fig:Imdeltas}}
\end{figure}

\paragraph{Limits from ${\epsilon}$}
Varying the mass of the gluino for fixed values of $m_{\tilde q}$, we found the upper limits on
$\sqrt{|\text{Im}((\delta^d_{XY})_{12} (\delta^d_{X'Y'})_{12})|}$
that are presented in \Figref{fig:Imdeltas}.
As expected, the limits from $\epsilon$ prove to be indeed very strict. If only the parameter $\sqrt{|\text{Im}(\delta^d_{\text{LL}})^{2}_{12}|}$ is not zero, the limits, varying according the mass of the squarks, are of order $10^{-2}$.
If only $\sqrt{|\text{Im}(\delta^d_{\text{LR}})^2_{12}|}$ is not zero,
the limits decrease to $\mathcal{O}(10^{-3})$.
If two MI parameters are non-zero, the corresponding limit is of order $10^{-3}$ or less,
and always smaller than for only one non-vanishing parameter.
Again, the lower the mass $m_{\tilde q}$, the lower the limits.

\subsection{Scenarios with mixing of the third generation}

We have considered a scenario where the squarks of the first and second
generation do not mix with those of the third generation.  In this case
it is sufficient to consider diagrams with at most two MI\@.  If there
is mixing with the third-generation squarks, diagrams with three or four
MI can give relevant contributions, especially in the scenario with a
light third generation.  For example, diagrams of the type (IIb) in
\Figref{fig:effdiagrams} yield a contribution proportional to
$(\delta^d_{XY})_{12} (\delta^d_{XY})_{13} (\delta^d_{XY})_{23}$.
However, if the third generation is light, these diagrams involve two
very different squark mass scales.  Already at the LO all momenta
between these scales enter, requiring a more complicated treatment of
the evolution from $m_{\tilde q}$ to $m_{\tilde g}$
\cite{Barbieri:2010ar,Bertuzzo:2010un}.  Taking into account these
subtleties, bounds on the heavy squark masses were derived in a scenario
with effective Minimal Flavour Violation and mixing between all squark
families in \cite{Barbieri:2010ar}.

Scenarios constrained by a family symmetry like the one of
\cite{Giudice:2008uk} allow for mixing with the third squark generation
but forbid mixing between the first and the second generation.  Thus,
they again avoid the complications due to diagrams containing both heavy
and light squarks and allow to derive limits on the MI parameters
$(\delta^d_{XY})_{13}$ and $(\delta^d_{XY})_{23}$, which are exactly
complementary to our bounds on $(\delta^d_{XY})_{12}$.  Such constraints
were calculated for the case of a light third generation in
\cite{Mescia:2012fg}, confirming that even after taking into account
additional observables from $B$ physics, $\epsilon$ still yields the
most severe constraints on some parameters.

So far, no constraints are available on scenarios with three heavy
squark generations and simultaneous mixing between all three families.

\section{Conclusions}
Supersymmetric scenarios involving light gluinos  and heavy squarks
have lately regained a considerable amount of attention,
since they may be a way to realize supersymmetry in agreement with the
LHC exclusion limits and with an acceptable amount of tuning.  As it is
also well-known, in such scenarios dangerous FCNC can occur especially
in the kaon sector. In an earlier work \cite{Kadota:2011cr}, we
considered a case with three heavy families of squarks with similar
masses above $20\TeV$. In order to estimate the gluino contributions to
$\Delta m_K$ and the CP-violating parameter $\epsilon$, we used the
evolution of Wilson coefficients from reference \cite{Contino:1998nw},
where a scenario with two heavy families of squarks and a lighter third
family with a mass comparable to that of the gluino was considered.  
In order to quantify precisely the accuracy of this estimate, we have
reviewed the way NLO QCD corrections are addressed in supersymmetric
scenarios involving heavy squarks and light gluinos.  We have calculated
the renormalization group evolution of the Wilson coefficients of the
$\Delta S = 2$ operators at NLO for scenarios with two and three heavy
squark families. This is codified in \eq{eq:Magic}.    

The SM determinations of $\Delta m_K$ and $\epsilon$ have improved significantly over the last decade. Together with important improvements on the lattice QCD bag parameters entering the effective Hamiltonian of the $\Delta S=2$ transitions, this proves relevant for re-assessing the limits on supersymmetric scenarios contributing to  $\Delta m_K$ and $\epsilon$. This was the second motivation for this work.

We have then determined the lower limits on the mass of the heavy
squarks, $m_{\tilde q}$, coming from $\Delta m_K$, using the MI
approximation with a value of $\sqrt{|\text{Re}((\delta^d_{XY})^2_{12})|}=0.22$,
$X,Y \in \{$L,\,R$\}$, for one or two flavour-violating parameters.
For cases with only one non-zero flavour-violating parameter, the limits
on $m_{\tilde q}$ for a gluino mass of about $900\GeV$ and three heavy
squark families are as follows: if only either
$\sqrt{|\text{Re}((\delta^d_{\text{LL}})^2_{12})|}=0.22$ or 
$\sqrt{|\text{Re}((\delta^d_{\text{RR}})^2_{12})|}=0.22$ is
non-zero, $m_{\tilde q}$ has to be larger than $5.9\TeV$.
If only either
$\sqrt{|\text{Re}((\delta^d_{\text{LR}})^2_{12})|}=0.22$ or 
$\sqrt{|\text{Re}((\delta^d_{\text{RL}})^2_{12})|}=0.22$ is
non-zero, the limit is above $13\TeV$.
If only the first two squark families are heavy, the limit changes by
about $100\GeV$.  For the case of
$\sqrt{|\text{Re}( (\delta^d_\text{LL})_{12} (\delta^d_\text{RR})_{12}  ) |} =0.22$ and the case of  $\sqrt{|\text{Re}( (\delta^d_\text{LR})_{12} (\delta^d_\text{RL})_{12}  ) |} =0.22$, while the
other flavour-violating parameters are zero, the limits are respectively
around $62\TeV$ and $165\TeV$, again for gluinos around $900\GeV$.  For
two heavy families of squarks, the limits are a few TeV larger than for
three heavy families.  We conclude that when the gluino contribution to
$\Delta m_K$ is close to its upper limit, it changes only by a few
percent if the mass of the third-generation squarks varies between
$m_{\tilde g}$ and $m_{\tilde q}$.

We have also noticed that indeed the NLO QCD evolution, as opposed to
just the LO evolution, proves relevant, especially for cases where there
is more than one type of flavour-violating parameter
$(\delta^d_{XY})_{12}$ involved \cite{Contino:1998nw}. As we have seen,
the difference can be as large as 20\,\%.

Finally, we have obtained bounds on the real and imaginary 
parts of the combinations
of flavour-violating parameters $(\delta^d_{XY})_{12}$ that
enter into the Wilson coefficients $C^{\tilde g}_i$ and $\tilde C^{\tilde g}_i$
for three heavy families of squarks.  
The former bounds stem from $\Delta m_K$ and the
latter from $\epsilon$.  We have explored these limits for squark masses
between 4 and $10\TeV$, varying the mass of the gluino for each value.
In short, when only the
real part of one type of parameter $(\delta^d_{XY})_{12}$ is
allowed to exist and for masses of the gluino
above $900\GeV$, the limit on 
$|\text{Re}(\delta^d_{XY})_{12}|$ is at
most of $\mathcal{O}(10^{-1})$ in most cases, of course the lower the
mass $m_{\tilde q}$, the lower the limit.  When only 
$\sqrt{|\text{Im} (\delta^d_{XY})^2_{12} |}$ for one type of parameter
is allowed to exist, its limit
is of $\mathcal{O}(10^{-2})$
or smaller.  Of course, when several types of flavour- and CP-violating
parameters $(\delta^d_{XY})_{12}$ are non-vanishing, the upper limits
become smaller.

\section*{Acknowledgments}
We would like to thank Joachim Brod, Christian Hoelbling, Luca Silvestrini and Javier Virto for very helpful discussions. 
This work was supported by the German Research Foundation (DFG) via the
Junior Research Group ``SUSY Phenomenology'' within the Collaborative
Research Center 676 ``Particles, Strings and the Early Universe'' and by
the INFN\@.
We acknowledge the Aspen Center for Theoretical Physics for a very stimulating environment which prompted the beginning of this work.  L.~V-S thanks the  University of Hamburg for its hospitality. Finally, we thank the Galileo Galilei Institute for Theoretical Physics for its hospitality during later stages of the
work.

\appendix

\section{Experimental information} \label{app:exp}
Values of experimental and lattice QCD parameters not given in the main
part are listed in \Tabref{tbl:expinputs}.  We are aware that there have
recently been some improvements in the determination of $B_K$
\cite{Durr:2011ap,Gamiz:2006sq,Laiho:2011dy,Kelly:2012uy,Bae:2011ff,Hoelbling:2012rj},
which have been averaged to a value of $0.7643\pm 0.0097$
\cite{Laiho:2012ss}.  {Adopting this average would change the values of the SM
predictions} for $\Delta m_K$ and $\epsilon$ obtained in \cite{Brod:2011ty}, but we have checked that the impact on our results
is negligible, since the limits are mainly determined by
$\sigma_{\Delta m_K}$ and $\sigma_{\epsilon}$.

We use the values of the bag parameters $B_i$ given by the ETM
collaboration \cite{Bertone:2012cu} because they provide the results in
the RI scheme at $2\GeV$.  {The RBC and UKQCD collaborations recently
also reported new computations of the relevant matrix
elements~\cite{Boyle:2012qb}, which are in good agreement with the ETM
calculations.  Thus, we estimate that employing the results of
\cite{Boyle:2012qb} would change our limits by less than $10\,\%$.}
\begin{table}
\label{tbl:expinputs}
\centering
\begin{tabular}{|l|l|l|}
\hline
\multicolumn{3}{c}{Experimental \& lattice inputs} \\ \hline
$\alpha_s(M_Z)$      & $0.1184\pm 0.0007$   & \cite{Beringer:2012}\\
$m_t := M_t^{\rm{Pole}}$  & $(173.5\pm 0.6\pm 0.8) \GeV$&  \cite{Beringer:2012}\\
$m_b := m_b(m_b)$   & $(4.18\pm 0.03) \GeV$ &  \cite{Beringer:2012}\\
$f_K$                          & $(0.1561 \pm 0.00085) \GeV$        & \cite{Beringer:2012}\\
$M_K$                        & $(0.497614 \pm 0.000022) \GeV$ &\cite{Beringer:2012}\\
$B_1$ & $0.51\pm0.02 $ &  \cite{Bertone:2012cu} \\
$B_2$ & $0.73\pm 0.04$  &  \cite{Bertone:2012cu}\\
$B_3$ & $1.29\pm 0.11 $ &  \cite{Bertone:2012cu}\\
$B_4$ & $1.04\pm 0.07 $ &\cite{Bertone:2012cu}\\
$B_5$ & $0.76\pm 0.09$ &  \cite{Bertone:2012cu}\\
$m_s(2\GeV)^\text{RI}$ & $(0.12\pm 0.006) \GeV$ & $^*$ \\
$m_d(2\GeV)^\text{RI}$ & $(0.006\pm 0.001) \GeV$ & $^*$ \\
\hline
\end{tabular}
\caption{Experimental and lattice QCD values used for our
analysis.  The bag parameters $B_i$ are given in the RI scheme
{at $2\GeV$}. $^*$We have computed with {\it RunDec} \cite{Chetyrkin:2000yt} the values of $m_s$ and $m_d$ in the RI scheme, using as input the values given by the PDG \cite{Beringer:2012} in the $\msbar$ scheme.}
\end{table}

\section{Beta functions} \label{app:BetaFunctions}

The two-loop $\beta$ functions
\begin{equation}
	\beta_X = \mu \frac{dX}{d\mu} = \beta_X^{(1)} + \beta_X^{(2)}
\end{equation}
for the strong gauge coupling and the gluino mass for the case of three
heavy squark families are equal to those of the  Split SUSY scenario.
In the limit of pure QCD, they read \cite{Tamarit:2012ie}
\begin{eqnarray}
\label{eq:betafuncts_3heavysquarks}
\beta^{(1)}_{g_s}&=&\frac{-1}{16 \pi ^2} \, 5 g_s^3 ,\nonumber\\
\beta^{(2)}_{g_s}&=&\frac{1}{(16\pi^2)^2} \, 22 g_s^5 ,\nonumber\\
\beta^{(1)}_{M_3}&=&\frac{-M_3}{16 \pi ^2}  \, 18 g_s^2 ,\nonumber \\
\beta^{(2)}_{M_3}&=&\frac{-M_3}{(16\pi^2)^2} \, \frac{429}{2} g_s^4 .
\end{eqnarray}

For the case of one family of light sfermions and two heavy ones, we
took the QCD limit of the two-loop $\beta$ functions of the
Effective SUSY scenario \cite{Tamarit:2012ie}.  Due to the breaking of
SUSY in the effective theory below $m_{\tilde q}$, the
gluino-squark-quark couplings and the squark quartic couplings are no
longer given by the gauge coupling.  In principle, these couplings are
also different for different squarks.  However, as we consider only QCD,
there is a single gluino-squark-quark coupling $\hat g_s$ and a single
quartic coupling $\gamma_s$, which are related to the gauge coupling at
$m_{\tilde q}$ by Eqs.~\eqref{eq:matchingmsq}.  The $\beta$ functions
relevant for our calculation are
\begin{eqnarray}
\label{eq:betafuncts_2heavysquarks}
\beta^{(1)}_{g_s}&=&\frac{1}{16 \pi ^2}\left(-\frac{13}{3} g_s^3\right),\nonumber\\
\beta^{(2)}_{g_s}&=&\frac{1}{(16\pi^2)^2}\left(\frac{110}{3}g_s^5-\frac{26}{3} g_s^3 \hat g_s^2\right),\nonumber\\
\beta^{(1)}_{\hat g_s}&=&\frac{1}{16 \pi ^2}\left( 3 \hat g_s^3-13 g_s^2 \hat g_s \right),\nonumber\\
\beta^{(2)}_{\hat g_s}&=&\frac{1}{(16\pi^2)^2}\left(-\frac{61}{9}\hat g_s^5 + \frac{125}{3} g_s^2 \hat g_s^3 - \frac{844}{9} g_s^4 \hat g_s 
-\frac{55}{18}  \hat g_s^3 \gamma_s
+\frac{11}{9} \hat  g_s \gamma_s^2
\right),\nonumber\\
\beta^{(1)}_{M_3}&=&\frac{M_3}{16 \pi ^2}
\left(-18 g_s^2 + 2 \hat g_s^2\right),\nonumber\\
\beta^{(2)}_{M_3}&=&\frac{M_3}{(16\pi^2)^2}\left( -206 g_s^4 - \frac{1966}{3} g_s^2 \hat g_s^2 - \frac{16}{3} \hat g_s^4 \right),\nonumber\\
\beta^{(1)}_{\gamma_s}&=& \frac{1}{16\pi^2}\left( 5 \gamma_s^2 - 16 g_s^2 \gamma_s + \frac{16}{3} \hat g_s^2 \gamma_s + 5 g_s^4 - \frac{14}{3} \hat g_s^4 \right).
\end{eqnarray}
The quartic coupling enters only via the two-loop part of $\beta_{\hat g_s}$.
Hence, it is sufficient to consider its one-loop running in order to
determine the two-loop running of $g_s$ and $M_3$.

As we have mentioned in section \ref{ssec:3heavysq}, our inputs are
the gluino pole mass $m_{\tilde g}$ and the scale $m_{\tilde q}$ at
which the heavy squarks decouple.  To begin with, we determine the value of
$\alpha_s(m_{\tilde g})$ from the experimental value at $M_Z$ and the SM
$\beta$ function.  Then we convert $m_{\tilde g}$ to the $\msbar$
running mass $M_3(m_{\tilde g})$ via \eq{eq:GluinoMassConversion}.  Initially guessing values for
$\hat g_s(m_{\tilde g})$ and $\gamma_s(m_{\tilde g})$, we run up to the
scale $m_{\tilde q}$, where we apply the $\msbar$ matching conditions
\cite{Tamarit:2012ry}
\begin{eqnarray}
\label{eq:matchingmsq}
\hat g_s(m_{\tilde q}) &=&
\sqrt{2} g_s(m_{\tilde q}) \left( 1 + \frac{g_s(m_{\tilde q})^2}{12\pi^2} \right) ,
\nonumber\\
\gamma_s(m_{\tilde q}) &=& g_s(m_{\tilde q})^2 .
\end{eqnarray}
We then run down to $m_{\tilde g}$ and again set the QCD coupling to the correct value determined initially. This procedure is repeated until we obtain both the right value of $\alpha_s(m_{\tilde g})$ and couplings satisfying the matching conditions \eqref{eq:matchingmsq}.

\section{Magic numbers \label{eq:magicnumbers}}
\subsection{Evolution below $m_{\tilde g}$}

\begin{equation}
a_r = \left( \frac{2}{7},\, -\frac{8}{7},\, \frac{1}{7},\,
 \frac{1-\sqrt{241}}{21},\, \frac{1+\sqrt{241}}{21} \right)_r \approx
(0.29,\, -1.1,\, 0.14,\, -0.69,\, 0.79)_r
\end{equation}

\begin{eqnarray}
b_{\text{LO}11}^{(r)} &=& (0.77,0,0,0,0)_r \nonumber\\
b_{\text{LO}22}^{(r)} &=& (0,0,0,1.8,0.0083)_r \nonumber\\
b_{\text{LO}23}^{(r)} &=& (0,0,0,-0.48,0.13)_r \nonumber\\
b_{\text{LO}32}^{(r)} &=& (0,0,0,-0.12,0.032)_r \nonumber\\
b_{\text{LO}33}^{(r)} &=& (0,0,0,0.032,0.48)_r \nonumber\\
b_{\text{LO}44}^{(r)} &=& (0,2.8,0,0,0)_r \nonumber\\
b_{\text{LO}45}^{(r)} &=& (0,0.94,-0.29,0,0)_r \nonumber\\
b_{\text{LO}55}^{(r)} &=& (0,0,0.88,0,0)_r \\
b_{\text{NLO}11}^{(r)} &=& (0.045,0,0,0,0)_r \nonumber\\
b_{\text{NLO}22}^{(r)} &=& (0,0,0,0.51,0.0020)_r \nonumber\\
b_{\text{NLO}23}^{(r)} &=& (0,0,0,-0.13,0.030)_r \nonumber\\
b_{\text{NLO}32}^{(r)} &=& (0,0,0,0.088,-0.0036)_r \nonumber\\
b_{\text{NLO}33}^{(r)} &=& (0,0,0,-0.023,-0.055)_r \nonumber\\
b_{\text{NLO}44}^{(r)} &=& (0,1.3,0,0,0)_r \nonumber\\
b_{\text{NLO}45}^{(r)} &=& (0,0.42,0.096,0,0)_r \nonumber\\
b_{\text{NLO}54}^{(r)} &=& (0,0.14,0,0,0)_r \nonumber\\
b_{\text{NLO}55}^{(r)} &=& (0,0.048,-0.063,0,0)_r \\
c_{11}^{(r)} &=& (-0.015,0,0,0,0)_r \nonumber\\
c_{22}^{(r)} &=& (0,0,0,-0.19,-0.0026)_r \nonumber\\
c_{23}^{(r)} &=& (0,0,0,-0.015,0.0063)_r \nonumber\\
c_{32}^{(r)} &=& (0,0,0,0.012,-0.0099)_r \nonumber\\
c_{33}^{(r)} &=& (0,0,0,0.00096,0.024)_r \nonumber\\
c_{44}^{(r)} &=& (0,-0.51,0.0054,0,0)_r \nonumber\\
c_{45}^{(r)} &=& (0,-0.26,-0.0061,0,0)_r \nonumber\\
c_{54}^{(r)} &=& (0,0,-0.016,0,0)_r \nonumber\\
c_{55}^{(r)} &=& (0,0,0.018,0,0)_r
\end{eqnarray}

\subsection{Evolution above $m_{\tilde g}$}

\begin{eqnarray}
d_{11}^{(r)} &=& (1.0,0,0,0,0)_r \nonumber\\
d_{22}^{(r)} &=& (0,0,0,0.98,0.017)_r \nonumber\\
d_{23}^{(r)} &=& (0,0,0,-0.26,0.26)_r \nonumber\\
d_{32}^{(r)} &=& (0,0,0,-0.064,0.064)_r \nonumber\\
d_{33}^{(r)} &=& (0,0,0,0.017,0.98)_r \nonumber\\
d_{44}^{(r)} &=& (0,1.0,0,0,0)_r \nonumber\\
d_{45}^{(r)} &=& (0,0.33,-0.33,0,0)_r \nonumber\\
d_{55}^{(r)} &=& (0,0,1.0,0,0)_r
\end{eqnarray}
These magic numbers correspond to LO effects.  Consequently, they are
independent of $\tilde n_f$, in particular of the number of heavy squarks.

\subsubsection{Three heavy squark generations\label{sbsc:magicnumbers3H}}

\begin{equation}
a'_r = \left( \frac{2}{5},\, -\frac{8}{5},\, \frac{1}{5},\,
 \frac{1-\sqrt{241}}{15},\, \frac{1+\sqrt{241}}{15} \right)_r \approx
(0.40,\, -1.6,\, 0.20,\, -0.97,\, 1.1)_r
\end{equation}

\begin{eqnarray}
e_{11}^{(r)} &=& (0.0019,0,0,0,0)_r \nonumber\\
e_{22}^{(r)} &=& (0,0,0,0.15,0.00044)_r \nonumber\\
e_{23}^{(r)} &=& (0,0,0,-0.040,0.0067)_r \nonumber\\
e_{32}^{(r)} &=& (0,0,0,0.013,-0.0059)_r \nonumber\\
e_{33}^{(r)} &=& (0,0,0,-0.0034,-0.091)_r \nonumber\\
e_{44}^{(r)} &=& (0,0.27,0,0,0)_r \nonumber\\
e_{45}^{(r)} &=& (0,0.091,0.037,0,0)_r \nonumber\\
e_{54}^{(r)} &=& (0,0.018,0,0,0)_r \nonumber\\
e_{55}^{(r)} &=& (0,0.0061,-0.035,0,0)_r \\
f_{11}^{(r)} &=& (-0.0019,0,0,0,0)_r \nonumber\\
f_{22}^{(r)} &=& (0,0,0,-0.15,-0.0044)_r \nonumber\\
f_{23}^{(r)} &=& (0,0,0,0.0088,0.025)_r \nonumber\\
f_{32}^{(r)} &=& (0,0,0,0.0098,-0.017)_r \nonumber\\
f_{33}^{(r)} &=& (0,0,0,-0.00058,0.095)_r \nonumber\\
f_{44}^{(r)} &=& (0,-0.28,0.0061,0,0)_r \nonumber\\
f_{45}^{(r)} &=& (0,-0.12,-0.0098,0,0)_r \nonumber\\
f_{54}^{(r)} &=& (0,0,-0.018,0,0)_r \nonumber\\
f_{55}^{(r)} &=& (0,0,0.029,0,0)_r
\end{eqnarray}

\subsubsection{Two heavy squark generations\label{sbsc:magicnumbers2H}}

\begin{equation}
a'_r = \left( \frac{6}{13},\, -\frac{24}{13},\, \frac{3}{13},\,
 \frac{1-\sqrt{241}}{13},\, \frac{1+\sqrt{241}}{13} \right)_r \approx
(0.46,\, -1.8,\, 0.23,\, -1.1,\, 1.3)_r
\end{equation}

\begin{eqnarray}
e_{11}^{(r)} &=& (-0.012,0,0,0,0)_r \nonumber\\
e_{22}^{(r)} &=& (0,0,0,0.19,-0.00026)_r \nonumber\\
e_{23}^{(r)} &=& (0,0,0,-0.050,-0.0040)_r \nonumber\\
e_{32}^{(r)} &=& (0,0,0,0.010,-0.0084)_r \nonumber\\
e_{33}^{(r)} &=& (0,0,0,-0.0027,-0.13)_r \nonumber\\
e_{44}^{(r)} &=& (0,0.35,0,0,0)_r \nonumber\\
e_{45}^{(r)} &=& (0,0.12,0.038,0,0)_r \nonumber\\
e_{54}^{(r)} &=& (0,0.018,0,0,0)_r \nonumber\\
e_{55}^{(r)} &=& (0,0.0060,-0.042,0,0)_r \\
f_{11}^{(r)} &=& (0.012,0,0,0,0)_r \nonumber\\
f_{22}^{(r)} &=& (0,0,0,-0.19,-0.0038)_r \nonumber\\
f_{23}^{(r)} &=& (0,0,0,0.020,0.035)_r \nonumber\\
f_{32}^{(r)} &=& (0,0,0,0.012,-0.014)_r \nonumber\\
f_{33}^{(r)} &=& (0,0,0,-0.0013,0.13)_r \nonumber\\
f_{44}^{(r)} &=& (0,-0.35,0.0060,0,0)_r \nonumber\\
f_{45}^{(r)} &=& (0,-0.14,-0.012,0,0)_r \nonumber\\
f_{54}^{(r)} &=& (0,0,-0.018,0,0)_r \nonumber\\
f_{55}^{(r)} &=& (0,0,0.036,0,0)_r
\end{eqnarray}

\addcontentsline{toc}{section}{References} \frenchspacing

\bibliographystyle{utphys}
\bibliography{wilson}

\end{document}